\documentclass[12pt]{article}

\usepackage{epsf}
\usepackage[dvips]{graphicx}

\newcommand{\bd}{\begin{document}}
\newcommand{\ed}{\end{document}}
\newcommand{\bs}{\begin{slide}}
\newcommand{\es}{\end{slide}}
\newcommand{\bc}{\begin{center}}
\newcommand{\ec}{\end{center}}
\newcommand{\ctr}[1]{\begin{center}#1\end{center}}
\newcommand{\be}{\begin{eqnarray}}
\newcommand{\ee}{\end{eqnarray}}
\newcommand{\ba}{\begin{eqnarray}}
\newcommand{\ea}{\end{eqnarray}}
\newcommand{\eq}[1]{\begin{eqnarray}#1\end{eqnarray}}
\newcommand{\nn}{\nonumber}

\begin{document}

\begin{center}
{ \bfseries
Renormdynamics, multiparticle production, negative binomial distribution and Riemann zeta function
}

\vskip 5mm

N.V. Makhaldiani

\vskip 5mm

{\small
{\it
Joint Institute for Nuclear Research\\
 Dubna, Moscow Region, Russia\\
}
{\it E-mail: mnv@jinr.ru
}}
\end{center}

\vskip 5mm

\begin{center}
\begin{minipage}{150mm}
\centerline{\bf Abstract}

Renormdynamic equations of motion
and their solutions are given. New equation for NBD distribution and Riemann zeta function invented.
Explicit forms of the
z-Scaling
functions are constructed.

\end{minipage}
\end{center}

\vskip 10mm

\section{Renormdynamics}
In quantum field theory (QFT) \cite{BSh} existence of a given theory means,
that we can control its
behavior at some scales (short or large distances) by
renormalization theory \cite{BSh,Collins}. If the theory exists, than
we want to solve it, which means to determine what happens on
other (large or short) scales. This is the problem (and content)
of Renormdynamics. The result of the Renormdynamics, the solution
of its discrete or continual motion equations, is the effective QFT
on a given scale (different from the initial one). We can invent
scale variable $\lambda$ and consider QFT on $D+1+1$ dimensional
space-time-scale. For the scale variable $\lambda\in (0,1]$ it is
natural to consider $q$-discretization, $0<q<1,\ {\lambda}_n=q^n,\
n=0, 1, 2,$ ... and $p$ - adic, nonarchimedian metric, with
$q^{-1}=p$ - prime integer number. The field variable
$\varphi(x,t,\lambda)$ is complex function of the real, x, t, and
p - adic, $\lambda,$ variables. The solution of the UV
renormdynamic problem means, to find evolution from finite to
small scales with respect to the scale (time)
$\tau=\ln\lambda/\lambda_0\in(0,-\infty).$ Solution of the IR
renormdynamic problem means to find evolution from finite to the
large scales, $\tau=\ln\lambda/\lambda_0\in(0,\infty).$
This evolution is determined by Renormdynamic motion equations with
respect to the scale-time.

\subsection{Renormdynamics of QCD}
The Renormdynamic (RD) equations play an important role in our understanding of
Quantum Chromodynamics (QCD) and the strong interactions. The beta function
is the most prominent object for QCD RD equations.
The calculation of the one-loop $\beta$-function in QCD has lead to the
discovery of asymptotic freedom in this model and to the establishment of QCD as the theory
of strong interactions \cite{tHooft,Gross,Politzer}.

The MS-scheme \cite{tHooft2} belongs to the class of massless schemes where the $\beta$-function does not depend on
masses of the theory and the first two coefficients of the $\beta$-function are scheme-independent.

RD equation for the coupling constant of QCD belongs to the following class of equations,
\ba\label{rde}
\dot{a}=\beta_1a+\beta_2a^2+...=\sum_{n\geq1}\beta_na^n,
\ea
where, in the case of the QCD,
\ba
a=(\frac{g}{4\pi})^2,\ \beta_1=\frac{D-4}{2},\ \beta_2=\frac{2}{3}N_F-\frac{11}{3}N_C,...
\ea
The equation (\ref{rde}) can be reparametrized,
\ba\label{rep}
&&a(t)=f(A(t))=A+f_2A^2+...+f_nA^n+...=\sum_{n\geq 1}f_nA^n,\cr
&&\dot{A}=b_1A+b_2A^2+...=\sum_{n\geq 1}b_nA^n,\\
&&\dot{a}=\dot{A}f'(A)=(b_1A+b_2A^2+...)(1+2f_2A+...+nf_nA^{n-1}+...)
\cr
&&
=\beta_1(A+f_2A^2+...+f_nA^n+...)
+\beta_2(A^2+2f_2A^3+...)+...+\beta_n(A^n+nf_2A^{n+1}+...)+...\cr
&&=\beta_1A+(\beta_2+\beta_1f_2)A^2+(\beta_3+2\beta_2f_2+\beta_1f_3)A^3+
...+(\beta_n+(n-1)\beta_{n-1}f_2+...+\beta_1f_n)A^n+...\cr
&&=\sum_{n,n_1,n_2\geq 1}A^nb_{n_1}n_2f_{n_2}\delta_{n,n_1+n_2-1}
=\sum_{n,m\geq1;m_1,...,m_k\geq 0 }A^n\beta_mf_1^{m_1}...f_k^{m_k}f(n,m,m_1,...,m_k),\cr
&&f(n,m,m_1,...,m_k)=\frac{m!}{m_1!...m_k!}\delta_{n,m_1+2m_2+...+km_k}\delta_{m,m_1+m_2+...+m_k},
\ea
\ba\label{crit}
&&b_1=\beta_1,\
b_2=\beta_2+f_2\beta_1-2f_2b_1=\beta_2-f_2\beta_1,\cr
&&b_3=\beta_3+2f_2\beta_2+f_3\beta_1-2f_2b_2-3f_3b_1=\beta_3+2(f_2^2-f_3)\beta_1,\cr
&&b_4=\beta_4+3f_2\beta_3+f_2^2\beta_2+2f_3\beta_2-3f_4b_1-3f_3b_2-2f_2b_3,...\cr
&&b_n=\beta_n+...+\beta_1f_n-2f_2b_{n-1}-...-nf_nb_1,...
\ea
so, by reparametrization, beyond the critical dimension ($\beta_1\neq 0$) we can change any coefficient but $\beta_1.$
We can fix any higher coefficient with zero value, if we take
\ba
f_2=\frac{\beta_2}{\beta_1},\ f_3=\frac{\beta_3}{2\beta_1}+f_2^2,\ ...\ , f_n=\frac{\beta_n+...}{(n-1)\beta_1},...
\ea
In this case we have exact classical dynamics in the (external) space-time and simple scale dynamics,
\ba
&&g=(\frac{\mu}{\mu_0})^{\frac{D-4}{2}}g_0=e^{-\varepsilon \tau}g_0;
\varphi(\tau,t,x)=e^{-(D-2)/2\tau}\varphi_0(t,x),
\psi(\tau,t,x)=e^{-(D-1)/2\tau}\psi_0(t,x)
\ea
We will consider in applications also the case when only one of the higher coefficients is nonzero.

In the critical dimension of space-time $\beta_1=0$ and we can change by reparametrization any coefficient but $\beta_2$ and $\beta_3.$
From the relations (\ref{crit}),
we can define the minimal form of the RD equation
\ba\label{eqA}
\dot{A}=\beta_2A^2+\beta_3A^3,
\ea
e.g. $b_4=0$ when
\ba
f_3=\frac{\beta_4}{\beta_2}+\frac{\beta_3}{\beta_2}f_2+f_2^2,
\ea
$f_2$ remains arbitrary and we can take e.g.
$f_2=0.$


We can solve (\ref{eqA}) as implicit function,
\ba
u^{\beta_3/\beta_2}e^{-u}=ce^{\beta_2t},\ u=\frac{1}{A}+\frac{\beta_3}{\beta_2},
\ea
than, as in the noncritical case, explicit solution will be given by reparametrization representation (\ref{rep}).
If we know somehow the coefficients $\beta_n,$ e.g. for first several exact and for others asymptotic values (see e.g. \cite{Kazakov Shirkov}) than we can construct reparametrization function (\ref{rep}) and find the dynamics of the running coupling constant.

\subsection{Renormdynamic functions (RDF)}
We
call RDF   functions $g_n=f_n(t),$ which are solutions of the RD motion equations
\ba\label{rden}
\dot{g}_n=\beta_n(g),\ 1\leq n\leq N.
\ea
In the simplest case of one coupling constant, the function $g=f(t),$
is constant $g=g_c$ when $\beta(g_c)=0,$ or
is invertible (monotone). Indeed,
\ba
\dot{g}=f'(t)=f'(f^{-1}(g))=\beta(g).
\ea
Each monotone interval ends by UV and IR fixed points and describes
corresponding phase of the system.

In the lattice (gauge) theory approach to the renormdynamics (see, e.g. \cite{MakhaldianiCQFT}), recently running
coupling constant dynamics were calculated for $SU(2)$ Yang-Mills model \cite{Bogolubsky}. The result is in agreement with perturbative calculations at small scales; at an intermediate scale the coupling constant reaches its maximum$(\simeq 1.25)$; than decrease. So, at the maximum, we may have nontrivial zero of the $\beta-$function, which corresponds to the conformal invariance of the gluodynamics at this point. Beyond this point we have another phase, strong coupling  phase with decreasing coupling constant similar (identical?!) to the abelian (monopole?) theory.

If we approximate the form of the
curve near maximum as
\ba
a(t)=a_c-b|t-t_c|^n,
\ea
for the $\beta-$function we obtain
\ba
\dot{a}=\beta(a,t)=sign(t_c-t)bn(\frac{a_c-a}{b})^{\frac{n-1}{n}}.
\ea
Of course this is not usual $\beta-$function, function of $a$ only. It depends also on $t.$  For $t>t_c$ we have perturbative phase. For $n>1,\ \beta(a_c,t)=0.$

The fundamental quark and gluon degrees of freedom are the relevant ones at high
temperatures and/or densities. Since these degrees of freedom are confined in the low temperature and density
regime there must be a quark and/or gluon (de)confinement phase
transition.

It is difficult to describe the phase transition because  there is not known a local parameter which can be linked to confinement. We consider the fractal dimension of the hadronic/quark-gluon space as order parameter of (de)confinement phase transition. It has a value less than 3 in the abelian, hadronic, phase, and more than 3, in nonabelian, quark-gluon, phase.

\subsection{Hamiltonian extension of Renormdynamics}
The renormdynamic motion equations (\ref{rden})
can be presented as nonlinear part of a hamiltonian
system with linear part \ba \dot{\Psi}_n=-\frac{\partial
\beta_m}{\partial g_n}\Psi_m, \ea the hamiltonian and canonical
Poisson bracket are \cite{Makhaldiani NP},
\ba H=\sum_{n=1}^N \beta(g)_n\Psi_n,\ \{
g_n,\Psi_m\}=\delta_{nm} \ea

In this extended version, we can define optimal control theory approach \cite{Pontriagin}
to the unified field theories. We can start from the unified value of the coupling constant, e.g.
$\alpha^{-1}(M)=29.0...$ at the scale of unification $M,$ in the minimal supersymmetric extension of the standard model
(see e.g. \cite{Kazakov}), put the aim to reach the SM scale with values of the
coupling constants measured in experiments, and find optimal threshold corrections to the RD coefficients.

 Any Hamiltonian quantum (and classical)
system can be described by infinitely divisible distributions
because, in the functional integral formulation, we use the
following step \ba U(t)=e^{-itH}=(e^{-i\frac{t}{N}H})^N \ea

In the case of the
scalar field theory, e.g.
\ba
L(\varphi)=\frac{1}{2}\partial_{\mu}\varphi
\partial^{\mu}\varphi-\frac{m^2}{2}\varphi^2-\frac{g}{n}\varphi^n
=g^{\frac{2}{2-n}}(\frac{1}{2}\partial_{\mu}\phi
\partial^{\mu}\phi-\frac{m^2}{2}\phi^2-\frac{1}{n}\phi^n), \ea
 so, to the constituent field
 $
 \phi_N,$ when $n>2,$ corresponds higher value of
 the coupling constant,
 \ba
g_N=gN^{\frac{n-2}{2}}
 \ea
 For weak nonlinearity, $n=2+2\varepsilon,\
  g_N=g(1+\varepsilon \ln N+O(\varepsilon^2))$

 \section{Negative binomial distribution }
Negative binomial distribution  (NBD)
\ba
P(n)=\frac{\Gamma(r+n)}{\Gamma(r)n!}(1-p)^rp^n=\frac{\Gamma(r+n)}{\Gamma(r)n!}(\frac{r}{<n>+r})^r(\frac{<n>}{<n>+r})^n,\ \sum_{n\geq 0}P(n)=1,
\ea
provides a very good parametrization for multiplicity distributions in $e^+e^-$ annihilation; in deep inelastic lepton scattering; in proton-proton collisions; in proton-nucleus
scattering.
Hadronic collisions at high energies (LHC) lead to charged multiplicity distributions whose shapes are well fitted by
a single NBD
in fixed intervals of central (pseudo)rapidity $\eta$ \cite{ALICE}.

The generating function for NBD is \ba \label{NBDGF}
&&F(h)=(1+\frac{<n>}{r}(1-h))^{-r}=(\frac{r}{<n>+r})^r(1-ph))^{-r}=\sum_{n\geq0}P(n)h^n
\ea

An useful property of
NBD with parameters $<n>$ and $r$ is that it is
the distribution of a sum of $r$ independent random variables
with a Bose-Einstein distribution\footnote{A Bose-Einstein, or geometrical, distribution is a thermal distribution for single state systems.}
and mean $<n>/r,$
\ba\label{NBDT}
&&p(n)=\frac{1}{<n>+1}(\frac{<n>}{<n>+1})^n=(e^{\beta\hbar \omega/2}-e^{-\beta\hbar \omega/2})e^{-\beta\hbar \omega (n+1/2)},
\cr
&&
\sum np(n)=<n>=\frac{1}{e^{\beta\hbar \omega}-1},\ f(x)=\sum_nx^np(n)=(1+<n>(1-x))^{-1} \cr
&&T=\frac{\hbar \omega}{\ln(1+\frac{1}{<n>})}\simeq \hbar \omega <n>,\ <n>\gg 1.
\ea
Temperature defined in (\ref{NBDT}) gives an estimation of the Glukvar temperature when it radiates hadrons.
We see that universality of  NBD in hadron-production is similar to the universality of black body radiation.
The generating function of a sum of independent random variables is the product of their generating functions. Indeed, for
\ba
n=n_1+n_2+...+n_k,
\ea
with
$n_i$
independent of each other, the probability distribution of $n$ is
\ba
&&P(n)=\sum_{n_1,...,n_k}\delta(n-\sum n_i)p(n_1)...p(n_k),\
F(h)=\sum_nh^nP(n)=f(h)^k
\ea
 An incoherent superposition of
 $m$ emitters that have
a negative binomial distribution with parameters $r, <n> $ produces
NBD with parameters
$mr, m<n>$.
So,
\ba F(r,<n>)^m=F(mr,m<n>) \ea
We can put
this equation in the closed nonlocal form \ba \label{SLF}
Q_qF=F^q,
\ Q_q=q^D, \ \
D=\frac{rd}{dr}+\frac{<n>d}{d<n>}=\frac{x_1d}{dx_1}+\frac{x_2d}{dx_2}
\ea
Note that $F(x_1,x_2)$ may be any function of the type
\ba
F(x_1,x_2)=f(\frac{x_1}{x_2})^{x_2}=\varphi(\frac{x_1}{x_2})^{x_1},\ \varphi(x)=f(x)^{\frac{1}{x}}.
\ea
By construction we know the solution of the nice equation
(\ref{SLF}) as GF of NBD, F.
We obtain corresponding differential equations, if we consider
$q=1+\varepsilon,$ for small $\varepsilon,$ \ba
&&(D(D-1)...(D-m+1)-(lnF)^m)\Psi=0,\
(\frac{\Gamma(D+1)}{\Gamma(D+1-m)}-(\ln F)^m)\Psi=0,
\cr
&&
(D_m-\Phi^m)\Psi=0,\  m=1, 2, 3, ..., \
D_m=\frac{\Gamma(D+1)}{\Gamma(D+1-m)}, \Phi=\ln F,
\ea with the
solution $\Psi=F=\exp(\Phi).$

\subsection{NBD,  Poisson and Gauss distributions} Fore high values of
$x_2=r$ the NBD distribution reduces to the Poisson distribution
\ba &&F(x_1, x_2, h)=(1+\frac{x_1}{x_2}(1-h))^{-x_2}\Rightarrow
e^{-x_1(1-h)}=e^{-<n>}e^{h<n>}=\sum P(n)h^n,\cr
&&P(n)=e^{-<n>}\frac{<n>^n}{n!} \ea For the Poisson distribution
\ba &&\frac{d^2F(h)}{dh^2}|_{h=1}=<n(n-1)>=<n>^2,\cr
&&D^2=<n^2>-<n>^2=<n>. \ea In the case of NBD, we have the
following dispersion low \ba D^2=\frac{1}{r}<n>^2+<n>, \ea which
coincides withe previous expression for high values of $r.$

Poisson GF belongs to the class of the infinitely divisible
distributions, \ba F(h,<n>)=(F(h,<n>/k))^k \ea
 For high values of $<n>,$ the Poisson distribution reduces
to the Gauss distribution \ba \label{Poisson}
P(n)=e^{-<n>}\frac{<n>^n}{n!}=
\frac{1}{\sqrt{2\pi<n>}}\exp(-\frac{(n-<n>)^2}{2<n>}) \ea

Let us consider the values $q=n, n=1, 2, 3, ...$ and take sum of the
corresponding equations (\ref{SLF}), we find \ba
\zeta(-D)F=\frac{F}{1-F} \ea

Now we invent a Hamiltonian $H$ with
spectrum corresponding to the set of nontrivial zeros of the zeta function,
in correspondence with Riemann hypothesis, \ba
&&-D_n=\frac{n}{2}+iH_n,\ H_n=i(\frac{n}{2}+D_n), \
D_n=x_1\partial_1+x_2\partial_2+...+x_n\partial_n,\cr
&&H_n^{+}=H_n=\sum_{m=1}^n H_1(x_m),\
H_1(x)=i(\frac{1}{2}+x\partial_x)=-\frac{1}{2}(x\hat{p}+\hat{p}x),\ \hat{p}=-i\partial_x \ea The Hamiltonian $H=H_n$ is
hermitian, its spectrum is real. The case $n=1$ corresponds to the
Riemann hypothesis.
The case $n=2$ corresponds to NBD, \ba\label{ME}
\zeta(1+iH_2)F=\frac{F}{1-F},\
F(x_1, x_2; h)=(1+\frac{x_1}{x_2}(1-h))^{-x_2} \ea

Let us scale $x_2\rightarrow \lambda x_2$ and take $\lambda \rightarrow \infty$ in (\ref{ME}), we obtain
\ba\label{ME2}
&&\zeta(\frac{1}{2}+iH(x))e^{-(1-h)x}=\frac{1}{e^{(1-h)x}-1},\
H(x)=
i(\frac{1}{2}+x\partial_x)=-\frac{1}{2}(x\hat{p}+\hat{p}x),\cr
&&
H(x)\psi_E=E\psi_E,\
 \psi_E=cx^{-s},\ s=\frac{1}{2}+iE,\ c=1/\sqrt{2\pi},
\cr
&&
\int_0^\infty dx \psi_E(x)^\ast\psi_{E'}(x)=\delta(E-E'),
\ea
\ba\label{IF}
&&\zeta(-D)e^{-x}=\zeta(\frac{1}{2}+iH(x))e^{-x}=\frac{1}{e^{x}-1},\\
&&
\int_0^\infty dxx^{s-1}\zeta(\frac{1}{2}+iH(x))e^{-x}
=<x^{s-1}|\zeta(\frac{1}{2}+iH(x))e^{-x}>
=\int_0^\infty dxx^{s-1}\frac{1}{e^{x}-1}=\Gamma(s)\zeta(s),
\cr
&&<x^{s-1}|\zeta(\frac{1}{2}+iH(x))e^{-x}>=<\zeta(\frac{1}{2}-iH(x))x^{s-1}|e^{-x}>\cr
&&=\zeta(\frac{1}{2}-iE)<x^{s-1}|e^{-x}>=\zeta(\frac{1}{2}-iE)\Gamma(s),\
\zeta(\frac{1}{2}-iE)=\zeta(s).
\ea

A slightly different consideration is the following.
 If we rescale $x\rightarrow xy$ in (\ref{IF}), multiply by $y^{s-1}$ and integrate by $y,$ we obtain usual integral
formula for zeta-function
\ba
&&\zeta(-D)\int_0^\infty y^{s-1}e^{-xy}dy=\int_0^\infty dy\frac{y^{s-1}}{e^{xy}-1},\cr
&&\zeta(-D)x^{-s}\Gamma(s)=x^{-s}\int_0^\infty dy\frac{y^{s-1}}{e^{y}-1},\cr
&&\zeta(s)=\frac{1}{\Gamma(s)}\int_0^\infty dy\frac{y^{s-1}}{e^{y}-1}
\ea

\section{Renormdynamical formulation of z - Scaling}


In the z - Scaling
approach to the inclusive multiparticle
distributions
(MPD)
(see, e.g. \cite{Tokarev - Zborovsky}),
different inclusive distributions depending on the variables
$x_1,... x_n,$ are described by universal function $\Psi(z)$ of a
fractal variable $z=x_1^{-\alpha_1}...x_n^{-\alpha_n}.$
It is interesting to find a dynamical system which generates
these
distributions and describes corresponding MPD.

We can find a good
function if we know its derivative. Let us consider the following
RD like equation \ba
z\frac{d}{dz}\Psi=V(\Psi),\
\int_{\Psi(z_0)}^{\Psi(z)}\frac{dx}{V(x)}=\ln\frac{z}{z_0},\ V(\Psi)=-\beta\Psi(z)+\gamma{\Psi(z)}^{1+n}. \ea

Corresponding
solution for $\Psi$ is \ba \label{z-scaling function}
\Psi(z)=\Psi(z;\beta,\gamma,c,n)=(\frac{\gamma}{\beta}+cz^{n\beta})^{-\frac{1}{n}}
\ea
where the integration parameter $c$ is defined from the normalization condition on $\Psi$ as
\ba c^{\beta n} =(\frac{\beta}{\gamma})
^{\frac{\beta -1}{\beta n}} \frac{\beta n}{B(\frac{\beta-1}{\beta
n}, \frac{1}{\beta n})} \ea
\begin{figure} [htb]
{\includegraphics*[width=0.6\textwidth]
{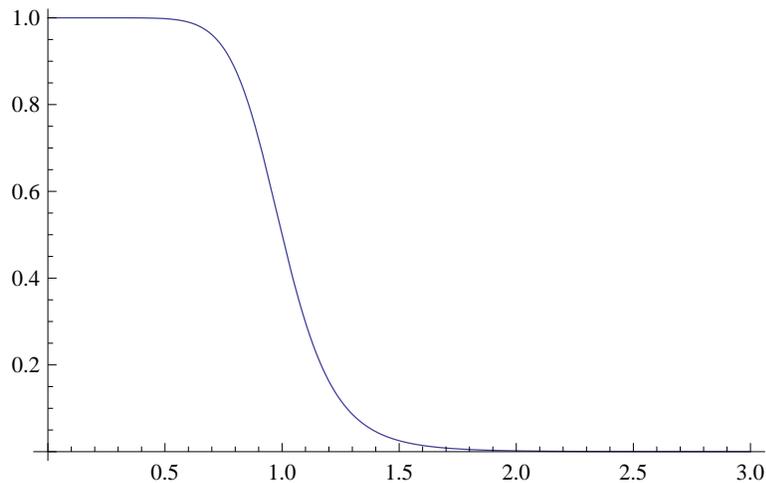}}
\centering
\caption{Typical form of z-scaling distribution (\ref{z-scaling function}),
$\Psi(z,9,9,1,1)$}
\label{fig:KNO}
\end{figure}

\section{Multiparticle production stochastic dynamics
} Let us imagine space-time
development of the the multiparticle process and try to describe
it by some (phenomenological) dynamical equation. We start to find the
equation for the Poisson distribution and than naturally extend
them for the NBD case.

Let us define an integer valued variable $n(t)$ as a number of
events (produced particles) at the time $t,\ n(0)=0.$ The
probability of event $n(t),\ P(t, n),$ is defined from the
following motion equation \ba \label{Poisson equation} &&P_t\equiv
\frac{\partial P(t, n)}{\partial t}=r(P(t, n-1)-P(t, n)),\
n\geq1\cr &&P_t(t, 0))=-rP(t, 0),\cr && P(t, n)=0,\ n<0,
 \ea
 so
 \ba
&&P(t, 0)\equiv P_0(t)=e^{-rt}, \cr &&P(t, n)=Q(t, n)P_0(t),\cr
&&Q_t(t, n)=rQ(t, n-1),\ Q(t, 0)=1.
 \ea
 To solve the equation for $Q,$ we invent its generating
 function
 \ba
F(t, h)=\sum_{n\geq0}h^nQ(t, n),
 \ea
 and solve corresponding equation
 \ba
F_t=rhF,\ F(t, h)=e^{rth}=\sum h^n\frac{(rt)^n}{n!},\ Q(t,
n)=\frac{(rt)^n}{n!},
 \ea
 so
 \ba
P(t, n)=e^{-rt}\frac{(rt)^n}{n!}
 \ea
 is the Poisson distribution.
 If we compare this distribution with (\ref{Poisson}), we identify
 $<n>=rt,$ as if we have a free particle motion with velocity $r$
 and the distance is the mean multiplicity. This way we have a connection between $n$-dimension of the multiplicity \cite{Baldin2008} and the usual dimension of trajectory.
 As the  equation gives right solution, its generalization may
 give more general distribution, so we will generalize the
 equation (\ref{Poisson equation}). For this, we put the equation in the closed form
 \ba &&P_t(t, n)=r(e^{-\partial_n}-1)P(t, n)\cr
 &&=\sum_{k\geq1}D_k\partial^kP(t, n),\ D_k=(-1)^k\frac{r}{k!},
 \ea
 where the $D_k,\ k\geq1,$ are generalized diffusion coefficients.

 For other values of the coefficients, we will have other
 distributions.
 \subsection{Fractal dimension of the multiparticle production trajectories  }
 For mean square deviation of the trajectory
 we have
 \ba
<(x-\bar{x})^2>=<x^2>-<x>^2\equiv D(x)^2\sim t^{2/d_f},
 \ea
 where $d_f$ is fractal dimension. For smooth classical trajectory
 of particles we have $d_f=1;$ for free stochastic, Brownian,
 trajectory, all diffusion coefficients are zero but $D_2,$ we have $d_f=2.$
 In the case of Poisson process we have,
 \ba
D(n)^2=<n^2>-<n>^2\sim t,\ d_f=2.
 \ea
 In the case of the NBD and KNO distributions
 \ba D(n)^2\sim t^2,\ d_f=1.
 \ea
 As we have seen, rasing $k,$ KNO reduce to the Poisson, so we have
 a dimensional (phase) transition from the phase with dimension 1 to
 the phase with dimension 2. It is interesting, if somehow this phase transition is connected to the other phase transitions in strong interaction processes.

 For the Poisson distribution GF
 is solution of the following equation,
 \ba
\dot{F}=-r(1-h)F,
 \ea
 For the NBD corresponding equation is
 \ba
\dot{F}=\frac{-r(1-h)}{1+\frac{rt}{k}(1-h)}F=-R(t)F,\
R(t)=\frac{r(1-h)}{1+\frac{rt}{k}(1-h)}.
 \ea
 If we change the time variable as $t=T^{d_f}$, we reduce the
 dispersion low from general fractal to the NBD like case.
 Corresponding transformation for the evolution equation is
 \ba
F_T=-d_fT^{d_f-1}R(T^{d_F})F,
 \ea
 we ask that this equation coincides with NBD motion equation, and
 define rate function $R(T)$
 \ba
d_fT^{d_f-1}R(T^{d_F})=\frac{r(1-h)}{1+\frac{rT}{k}(1-h)}, \ea now
the following equation defines a production processes with fractal
dimension $d_F$ \ba F_t=-R(t)F,\
R(t)=\frac{r(1-h)}{d_Ft^{\frac{d_F-1}{d_F}}(1+\frac{rt^{1/d_F}}{k}(1-h))}
 \ea

It is a pleasure to thank Yu.M.Bystritskiy, M.V.Tokarev and the
members of the seminar $\odot NM\pi$ for stimulating discussions
and various help.


\end{document}